\documentclass[aip,graphicx]{revtex4-2}
\usepackage{hyperref}
\usepackage{xr}

\usepackage{amsmath}
\usepackage{cleveref}
\usepackage{color}
\usepackage{graphicx}
\usepackage{dcolumn}
\usepackage{bm}
\usepackage{ulem}

\begin{document}
	
\title{Current-induced enhancement of photo-response in graphene THz radiation detectors}

\author{K. Indykiewicz}
\affiliation{Faculty of Electronics, Photonics and Microsystems, Wroc\l{}aw University of Science and Technology, 50-372 Wroc\l{}aw, Poland}
\author{C. Bray}
\affiliation{CNRS/Laboratoire Charles Coulomb (L2C), 34095  Montpellier,  France}
\author{C. Consejo}
\affiliation{CNRS/Laboratoire Charles Coulomb (L2C), 34095  Montpellier,  France}
\author{F. Teppe}
\affiliation{CNRS/Laboratoire Charles Coulomb (L2C), 34095  Montpellier,  France}
\author{S. Danilov}
\affiliation{University of Regensburg, Faculty of Physics, D-93053 Regensburg, Germany}
\author{S.D. Ganichev}
\affiliation{University of Regensburg, Faculty of Physics, D-93053 Regensburg, Germany}
\affiliation{CENTERA Laboratories, Institute of High Pressure Physics, Polish Academy of Sciences PL-01-142 Warsaw, Poland}
\author{A. Yurgens}
\email[]{yurgens-at-chalmers.se}
\affiliation{Chalmers University of Technology, SE-412 96 G\"oteborg, Sweden}

\date{\today}

\begin{abstract}
Thermoelectric readout in a graphene THz radiation detector requires a \textit{p-n} junction across the graphene channel. Even without an intentional \textit{p-n} junction, two latent junctions can exist in the vicinity of the electrodes/antennas through the proximity to metal. In a symmetrical structure, these junctions are connected back-to-back and therefore counterbalance each other with regard to rectification of the ac signal. Because of the Peltier effect, a small dc current results in additional heating in one- and cooling in another \textit{p-n} junction thereby breaking the symmetry. The \textit{p-n} junctions then no longer cancel, resulting in a greatly enhanced rectified signal. This allows to simplify the design and effectively control the sensitivity of the THz-radiation detectors.
\end{abstract}

\maketitle

The graphene-based Terahertz (THz) detectors can be fast and sensitive devices in a wide frequency range.\cite{Vicarelli_2012, Review_Koppens2014}  There are several readout mechanisms in graphene detectors such as bolometric,\cite{quantum_dots} thermoelectric (TEP), \cite{TEP_2014} ballistic,\cite{Teppe_ballistic} based on noise thermometry,\cite{noise_thermometry} ratchet effects,\cite{olbrich_ratchet_2016,fateev_ratchet_2017} and electron-plasma waves,\cite{resonant,tomadin_theory_2021} also called Dyakonov-Shur (D-S) mechanism.\cite{Dyakonov_Shur, Dyakonov_Shur_PRL} Detectors with the TEP readout mechanism are simple, do not require electrical bias and therefore have no 1/$f$ noise, allow for scalable fabrication using CVD graphene, and have undemanding electrical contacts. High efficiency of such detectors stems from a large radiation-induced increase of the electronic temperature $T_e$ because of a weak electron-phonon (\textit{e-ph}) coupling in graphene\cite{Supercollision_cooling, e-cooling} and a large value of the Seebeck coefficient ($S\sim T_e/3\ \mathrm{\mu V/K}$).\cite{TEP1,TEP2}   

\begin{figure}[!ht]
	\begin{center}
		\includegraphics[width=0.7\textwidth]{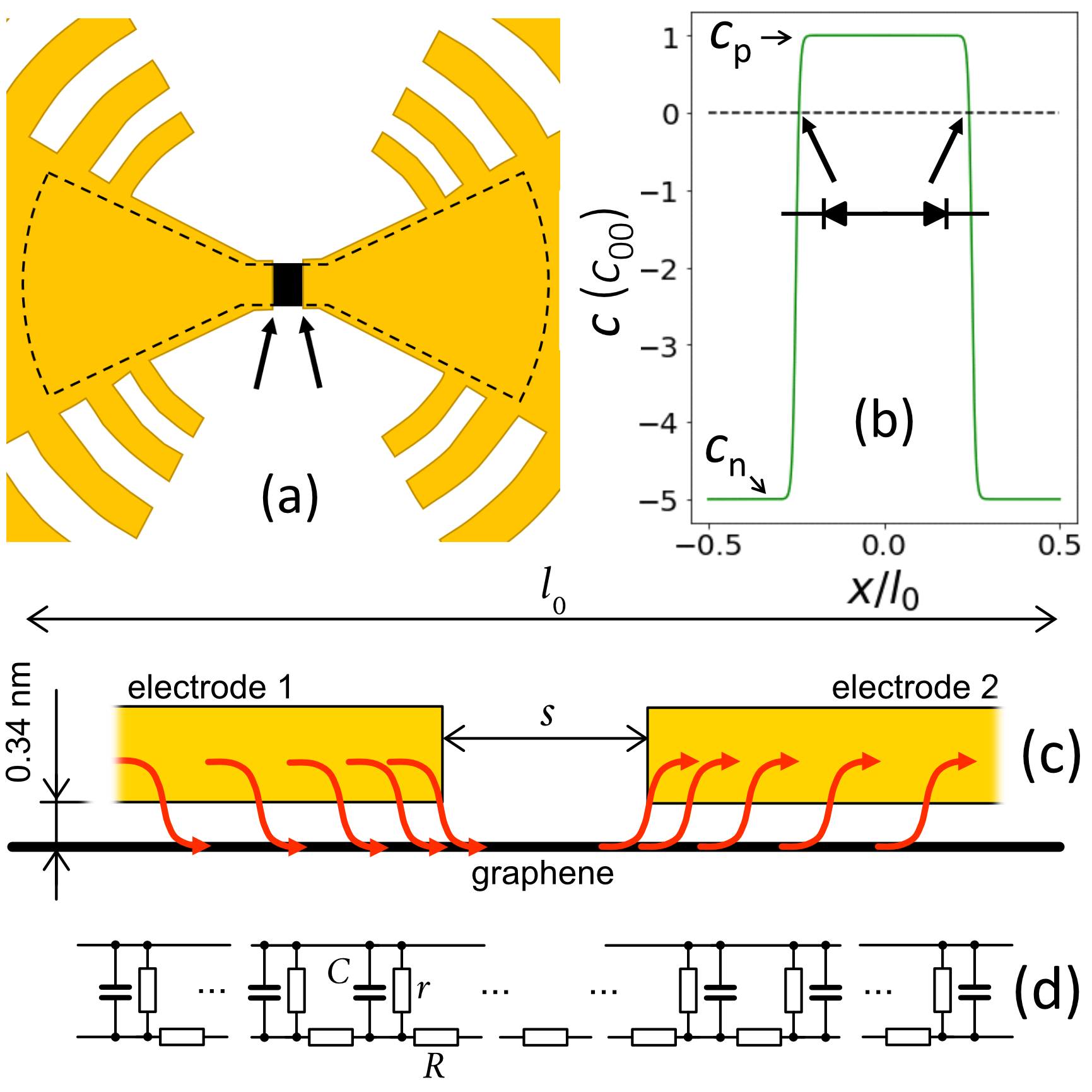}
		\vspace{-5mm}
	\end{center}
	\caption{(a) The model geometry of  a symmetric graphene detector.  Graphene is outlined by the dashed line.  The arrows mark two latent \textit{p-n} junctions in the vicinity of the electrodes (log-periodic antenna in this case). (b) Schematic doping profile in a device.  The regions under the electrodes are assumed to be $n$ doped because of the proximity to the metal. The latent \textit{p-n} junctions (diodes) are connected back to back (the inset). (c) Schematic cross section of graphene channel with two metal electrodes. Red arrows show a current flow and its distribution (crowding). (d) a lumped-element representation of the device, where $C$, $R$, and $G=1/r$ are the capacitance, graphene resistance, and contact conductance per unit length, respectively.  Note the similarity with the classical transmission line (see, e.g., \cite{carr_microwave_1996} or Wikipedia), allowing for a straightforward estimation of the current-crowding length $\lambda_j=1/\sqrt{R G} \sim 1-5\ \mathrm{\mu m}$. In the self-gating scenario, $R$ is a function of the local voltage drop $V_l$ across the contact resistance $r$, $R=\mu^{-1} (C^2V_l^2+c_{00}^2e_0^2)^{-0.5}$, which introduces a significant non-linearity at high bias.  Here, $c_{00}$ is the residual charge density and $e_0$ is electron charge.}
	\label{fig:geometry}
\end{figure}

A \textit{p-n} junction across the graphene channel must be formed to fully realize the TEP readout in a graphene-based radiation detector (see Fig.~\ref{fig:geometry}a). It can be done either chemically or electrostatically, by using a split top gate.\cite{Skoblin_APL} Without \textit{p-n} junctions, the TEP signal is usually insignificant.\cite{Vicarelli_2012, Review_Koppens2014}   

However, there can be latent \textit{p-n} (or \textit{p-p'} or \textit{n-n'}) junctions in the vicinity of the electrodes/antennas through the proximity to metal.\cite{doping-by-metal,cusati_2017,chaves_physical_2014} These junctions do not normally contribute to rectification~\footnote{Because of the ambipolarity of the transport properties, rectification in graphene \textit{p-n} junctions, i.e., the appearance of a dc signal stems from the thermoelectric effects.} of the ac current induced by THz radiation because the junctions (diodes) are connected back to back, i.e., symmetrically in the opposite directions (see Fig.~\ref{fig:geometry}b). Here, we show experimentally and by numerical simulations that a small dc current breaks the symmetry and the ac current gets rectified, which considerably increases the signal. This allows for an effective control of  sensitivity of the THz-radiation detector. 

We fabricated the devices from a chemically-vapor-deposited (CVD) graphene grown on a 2" large copper foil 25- or 60 $\mu$m thick in the commercial cold-wall CVD system (AIXTRON Black Magic II). Pure Ar and H$_2$ were used as a buffer- and nucleation-controlling gases, respectively. The precursor gas was CH$_4$ diluted in Ar (5\%). The nominal temperature was regulated by using a thermocouple in contact with the graphitic heater.  Many patches of two- and three layer graphene were seen in the majority of samples.  The resulting charge-carrier mobility $\mu$ of such a graphene transferred to ordinary office lamination foil (EVA/PET) was nonetheless surprisingly high, reaching $9000\ \mathrm{cm^2/(V s)}$.\cite{C7NR07369K, Munis}

The THz detectors were fabricated in many ways, with graphene both under- and on top of metal electrodes/antennas. We chose also different metals for the electrodes, Au, Pt, Pd, which were expected to have different proximity-doping effects on graphene.\cite{doping-by-metal} The CVD graphene was either transferred to SiO$_2$/Si substrate by using the PMMA- or paraffin-assisted technique,\cite{transfer-review} or simply glued to a substrate by an epoxy-based adhesive. Bow-tie or log-periodic antennas were lithographically patterned to have a better coupling to THz radiation (see Fig.~\ref{fig:geometry}a). However, the antennas appeared to only play a minor role in the frequency range of our measurements because of a relatively high graphene-to-metal contact resistance resulting in a significant impedance mismatch. This leaves spacey room for uncomplicated improvement of the detectors in the future, promising a much better performance than demonstrated in this work.  

For optical excitation, we used Gunn diodes and pulsed  THz laser\cite{ganichev1998,shalygin2007} optically pumped by a transversely excited atmospheric-pressure CO$_2$ laser.\cite{ganichev2003}  The Gunn diode provided a linearly polarized radiation  with the frequency of 94 GHz and estimated incident power from 1 to 10~mW. The radiation was modulated by an optical chopper at the frequency of 37~Hz, allowing measurements of photoresponse with the standard lock-in technique. The THz power delivered to the samples in the cryostat through the optical windows is somewhat difficult to reliably estimate because of the multiple reflections from the metal walls of the cryostat resulting in light interference and a complex pattern of maxima and minima of the light intensity.

The THz laser provides single pulses of monochromatic radiation with the pulse duration in the order of 100~ns, repetition rate of 1~Hz, and peak  power in the order of hundreds of kW.  The peak power was monitored with the THz photon-drag detectors.\cite{photon-drag} The laser operated at the frequencies $f = 0.61$, 1.07, 2.02, and 3.31~THz.  The photoresponse to the THz pulses was measured  with a digital oscilloscope as a voltage drop across 50-$\Omega$ load resistor. 

\begin{figure}[!ht]
	\begin{center}
		\includegraphics[width=0.8\textwidth]{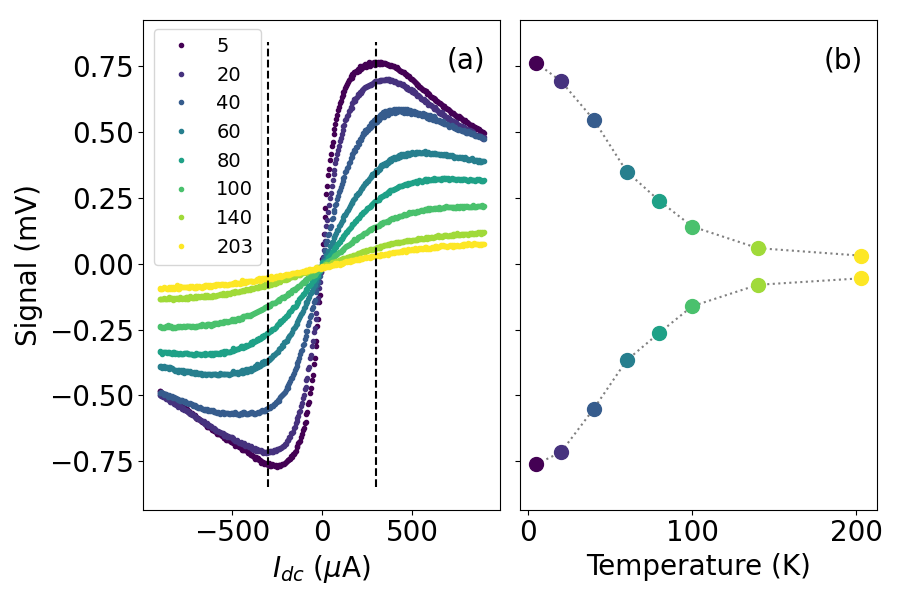}
		\vspace{-5mm}
	\end{center}
	\caption{Output response signal to 94-GHz radiation versus dc current (a) and the temperature dependence of the signal at the two fixed dc currents marked by the vertical dashed lines (b). }
	\label{fig:Al2O3}
\end{figure}

 Fig.~\ref{fig:Al2O3}a shows the response signal versus dc current demonstrating initially linear increase of the signal, which then have a tendency to saturation- and even decrease at the maximum current. The sign of the signal changes with the direction of dc current. In the samples with dissimilar metals on both ends of the graphene channel, there was usually an offset in the vertical direction common to all curves, which meant that the signal at low temperature was significant even at zero dc current.

The signal decreases with temperature (see Fig.~\ref{fig:Al2O3}b); the shape of this decreasing function is sample dependent. In Fig.~\ref{fig:Al2O3}b, the signal changes gradually and survives up to room temperature. However, in several other samples, the signal decayed to zero at $150-200$~K. A couple of devices showed a very abrupt change of the signal that vanished at already $\sim$40~K (see Supplementary material). The mechanism behind this temperature dependence is unclear and requires further experiments.

Since the response signal in our devices is due to the thermoelectric effects and involves electron heating, the decay of the signal with temperature should be largely attributed to increased cooling of hot electrons. The electrons are cooled by interactions with phonons. These interactions are generally weak because the population of optical phonons is exponentially small at low temperature. The cooling efficiency through the acoustical phonons is impeded because of the momenta mismatch, but can be somewhat improved when involving scattering by impurities (supercollisions).\cite{Supercollision_cooling}  However, there can be many other modes involving the out-of-plane direction in a multilayer graphene, e.g., the shear mode at 31~$cm^{-1}$.\cite{tan_shear_2012,ferrari_raman_2013} Many double-layer patches and these phonons can in principle be an effective channel for cooling of the electrons.   

The heating of electrons can be regarded by simply considering graphene as a conducting layer with a Drude-like frequency-dependent conductivity $\sigma(\omega)= \sigma_0(1-i\omega \tau)^{-1}$. The heating effects are described by the real part of the conductivity, $P(\omega)\sim v^2\mathrm{Re}(\sigma(\omega))\sim v^2\sigma_0/(\omega \tau)^2$, for $\omega \tau \geq 1$. Here, $P$ is the Joule heating power, $v$ is the ac-voltage amplitude in graphene, $\sigma_0$ is the dc conductivity, $\tau$ is the scattering time, $\omega =2\pi f$, and $f$ is the frequency. 

\begin{figure}[!ht]
	\begin{center}
		\includegraphics[width=0.5\textwidth]{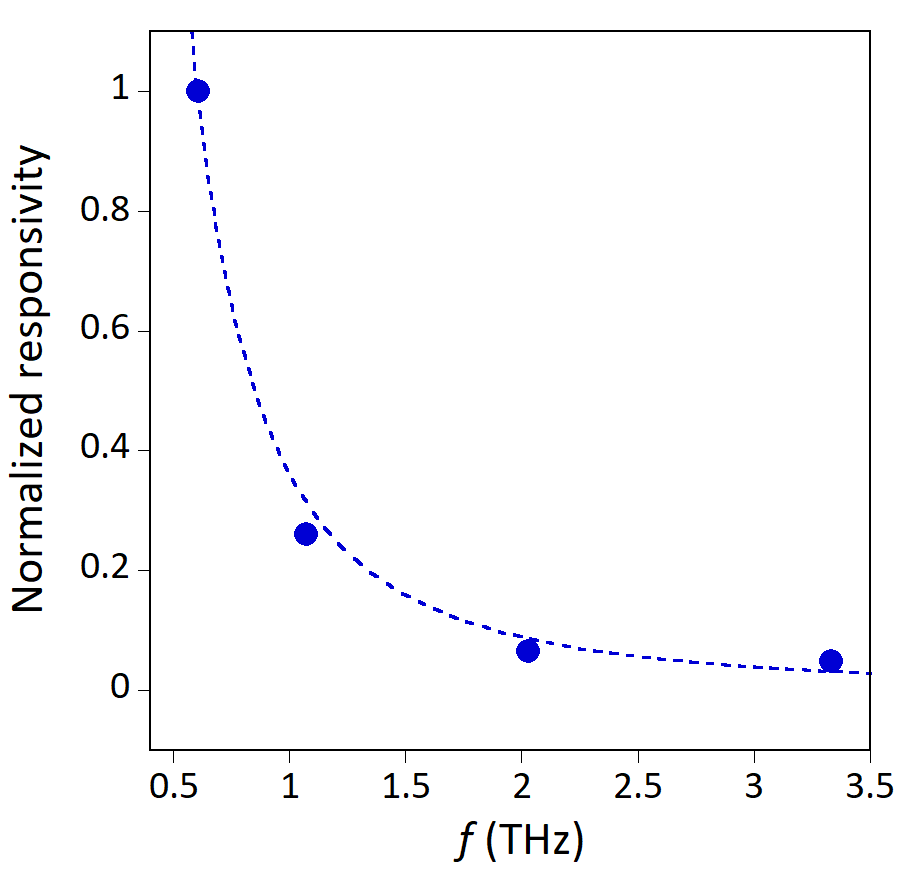}
		\vspace{-5mm}
	\end{center}
	\caption{The normalized responsivity $\hat{r}$ for the linear-polarized THz pulsed radiation at $T_0=300$~K.  Normalization constant is 0.072 $\mu$V/W. The dotted line is a fit to equation $\hat{r}=(f_0/f)^2$, with the parameter $f_0\approx 0.6$~THz.  }
	\label{fig:THz}
\end{figure}

Fig.~\ref{fig:THz} shows the frequency dependence of the normalized response signal at room temperature. Rotating the electric field vector in respect to the line connecting the contacts, we observed only a weak polarization dependence of the photoresponse, which confirms the anticipated inefficiency of the antennas because of the large impedance mismatch. The overall responsivity decays with the frequency as $1/f^2$ (see the dashed line in Fig.~\ref{fig:THz}), in correspondence with the suggested Drude model for Joule heating by the THz radiation.\cite{Drude_THz_2014}

\begin{figure}[!ht]
\begin{center}
 \includegraphics[width=0.8\textwidth]{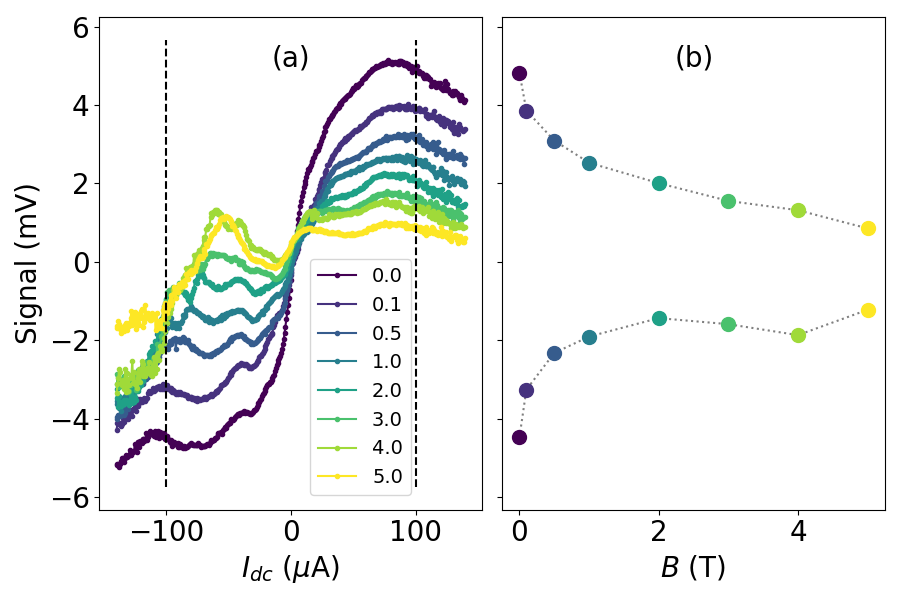}
 \vspace{-5mm}
\end{center}
  \caption{ (a) The response signal as a function of the dc current in different magnetic fields at 1.7~K. Note the overall decrease of the signal as the magnetic field increases. (b) The signal versus the magnetic field $B$ at two dc currents, indicated by the vertical dashed lines in (a).  }
  \label{fig:signal(B)}
\end{figure}

Applying perpendicular magnetic field to graphene, we observed that the photoresponse substantially decreased. This is shown in Fig.~\ref{fig:signal(B)}. Overall, the decrease of the signal in the magnetic field can be explained by an increased relaxation of electrons causing a decrease of their temperature and thermoelectric response. Indeed, it was experimentally observed in a graphene with defects that the relaxation of hot electrons would increase in the perpendicular magnetic field. This was attributed to the supercollision cooling combined with the presence of mirror-plane-symmetry-breaking defects in graphene.\cite{Ermin_supercollision}  It was suggested that the defect-electron interaction could activate the out-of-plane phonons at the $\Gamma$ point ($\Gamma$ZO phonons). Supercollisions involving these phonons would then allow transitions between the sufficiently broadened lowest Landau levels in realistic magnetic fields, thereby opening an additional channel for the electron-energy relaxation.\cite{Ermin_supercollision}

Many double-layer patches in our graphene devices could be the symmetry-breaking defects, validating this scenario. Moreover, the presence of soft shear-mode phonons (31~cm$^{-1}$) in multi-layer graphene can further emphasize the supercollision cooling in the magnetic field.\cite{tan_shear_2012} The shear mode energy roughly corresponds to 40~K when population of these phonons is expected to dramatically increase and provide an additional pathway for electron energy relaxation. This might be the reason for a sudden disappearance of the signal at this temperature in a couple of our samples (see Fig.~S2).

Some wiggles are also seen in the curves at 1.7 K (see Fig.~\ref{fig:signal(B)}). They are visible even at zero field but not in all samples (see e.g., Fig.~\ref{fig:Al2O3}) and become more pronounced with increasing magnetic field.  The structures are clearly seen only at negative dc currents. The nature behind these structures is not known and requires further studies.  Possible explanations include plasma resonances~\cite{resonant} or some microwave interference effects in the cavity formed between metallic sample holder and the electrodes, with graphene and dielectric substrate in between. The self-doping effect under the electrode can in principle explain the asymmetry with regard to the current direction in the latter case.    


The ac response resulting from the dc bias in the completely symmetric graphene-metal structures can be explained in two ways. One assumes the self-gating effect in the top contacts to graphene. Another takes into account the non-uniform doping in graphene because of proximity to metals. It can also be a combination of the two mechanisms in real devices. Both mechanisms are similar in that the Joule heating and thermoelectric effects give rise to a rectified signal because of a spatially non-uniform doping in graphene. The difference is in the manner of how the non-uniformity is created - through the self-gating or proximity doping (see Fig.~\ref{fig:geometry}). Here, we outline a general model of thermal balance between electron- and phonon  subsystems in graphene that is subject to Joule heating, in the presence of thermoelectric effects and non-uniform doping.


 A graphene strip of length $l_0$ and width $w$ is subdivided into $p$- and $n$ regions (see Fig.~\ref{fig:geometry}b). The $p$ region corresponds to graphene channel in between the source- and drain contacts, while $n$ regions - to graphene under the contacts because of proximity doping in graphene~\footnote{The sign of doping can vary depending on the metal.\cite{doping-by-metal} In most cases described here, it was Au/Ti contacts, which induce $n$ doping.}. 
 
 The strip rests on a SiO$_2$ layer 300~nm thick on top of Si substrate at the constant temperature $T_0$. The electrical current with the linear density $j$ flows in $x$ direction from the source- to drain electrodes.  Electrons in graphene are heated by the current and cooled by phonons through the electron-phonon interaction. Phonons escape into Si substrate via thermal resistance of the SiO$_2$ layer, which results in an increased lattice temperature $T_{ph}>T_0$. The heating- and temperature distribution $T_e(x)$ are highly non-uniform because the spatially varying doping profile $c(x)$ results in a non-uniform conductance and, hence, Joule heating. Moreover, it gives rise to a non-uniform Seebeck coefficient $S=S(x)$ and the Peltier effect, $(j T_e \partial S/\partial x)$. This model is described by the coupled one-dimensional heat-diffusion equations:
\begin{eqnarray}
-\frac{\partial}{\partial x} \Big( \kappa_e \frac{\partial T_e}{\partial x}\Big) &=&
\frac{j^2}{\sigma} -
j T_e \frac{\partial S}{\partial x} -
\alpha_i \big( T_e^i-T_{ph}^i \big)\label{eq:diffusion1}\\
- \kappa_{ph} \frac{\partial^2 T_{ph}}{\partial x^2} &=&
\alpha_i \big( T_e^i-T_{ph}^i \big) -
\kappa_0 \big( T_{ph}-T_0 \big) \label{eq:diffusion2}
\end{eqnarray}

\noindent where $\kappa_e = L_0 \sigma T$ is the electronic sheet thermal conductivity, $L_0$ is the Wiedemann–Franz constant, $T_e$, $T_{ph}$, $k_e$, and $k_{ph}$ are the temperatures and thermal conductivities of electronic ($e$) and phononic ($ph$) subsystems, respectively. $\kappa_0$ is the thermal conductivity of SiO$_2$ layer 300~nm thick. The Seebeck coefficient $S$ in graphene is assumed to obey Mott's equation. The heat transfer to the phonon system is described by the last term in Eq.~\ref{eq:diffusion1}. The exponent $i=3$ or $i=4$ at temperatures above or below the Bloch-Gr\"{u}neisen temperature $T_\mathrm{BG}$, respectively, and $\alpha_3 \propto c$.\cite{Supercollision_cooling}

Numerically solving Eq.~\ref{eq:diffusion1} and \ref{eq:diffusion2}  gives  $T_e(x)$, $T_{ph}(x)$, the TEP voltage, and the total Joule dissipation for any bias current $j$. The current in real detectors will be a sum of the dc- and ac components: $j(t)=j_{dc} + j_0 \sin(\omega t)$. For $\omega\ll 2\pi/\tau$, where $\tau<50$~fs is the electron-heating time \cite{Koppens_fs_nnano}, the responsivity can be found by averaging the TEP voltage and Joule power over one period of the ac bias. Simulation details can be found in the Supplementary material and Ref.\cite{me_Sensors}. 

\begin{figure}[!ht]
\begin{center}
 \includegraphics[width=0.8\textwidth]{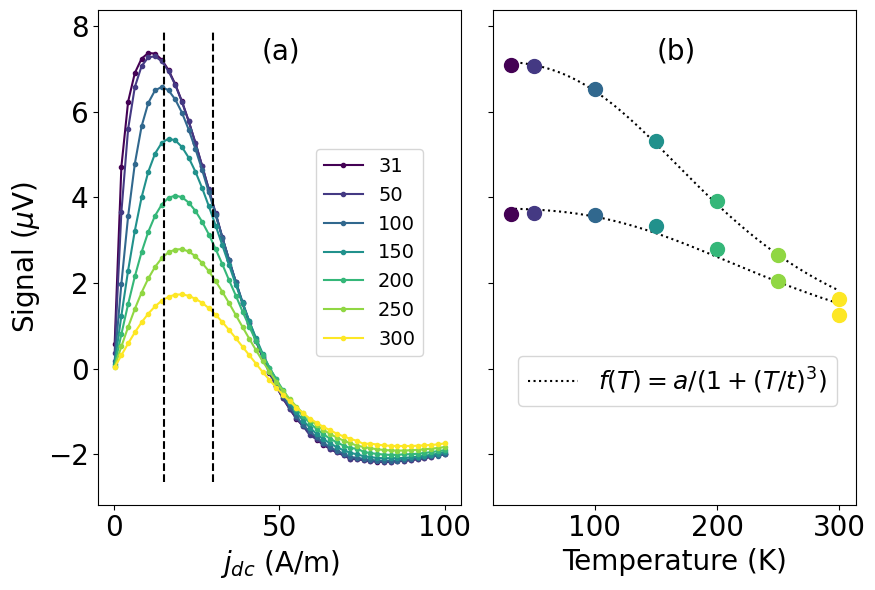}
 \vspace{-5mm}
\end{center}
  \caption{(a) The simulated response signal versus dc current at different temperatures. The following parameters were used: ($c_{00}$, $c_p$, $c_n$) = (0.5, 1, -5)$\times 10^{12}$~cm$^{-2}$, $\lambda_j=1\ \mathrm{\mu m}$, $j_{ac}=1$~A/m (see also Fig.~\ref{fig:geometry}b). (b) The temperature dependence of the maximum signal and the best fit to the empirical equation $f(T) = a/[1+(T/t)^3]$, with $t=210$ and 264~K for the top- and bottom curve, respectively. }
  \label{fig:static_n_signal}
\end{figure}

Simulations were conducted for various sets of parameters, giving somewhat different shapes of the resulting curves, depending on the parameters. In Fig.~\ref{fig:static_n_signal} it is seen, e.g., that the signal can even change sign as a function of dc current.  The phenomenological fit shown in Fig.~\ref{fig:static_n_signal}b has no real significance even though it reveals $T^{-3}$ factor, which much likely stems from the cooling term in Eq.~\ref{eq:diffusion1}. 

The main parameters that affect the shape of the curves are the doping- and residual charge densities. The signal amplitude is at maximum when there are \textit{p-n} junctions at the edges of the electrodes. In the case of \textit{n-n'} or \textit{p-p'} junctions, the signal is reduced. This is understandable from the fact that the graphene sheet resistance (and, hence, Joule heating) is at maximum where the doping is zero. The current crowding is beneficial because it also results in the Joule heating that takes place largely at the electrode edges. The maximum temperature increase in the \textit{p-n} junctions leads to the large thermoelectric voltage and increased signal as well.

In summary, we have shown that in symmetric graphene radiation detectors, the symmetry can be lifted by application of dc current. This leads to a non-compensated response of the detectors because of nonequivalent thermal conditions for the two \textit{p-n} junctions at the edges of metal electrodes. One \textit{p-n} junction is biased in the forward - while another - in the reverse direction, corresponding to the Peltier heating and cooling, respectively. The \textit{p-n} junctions can be formed because of proximity doping- or self-gating effect under the metal electrodes. The simulations reveal several possible scenarios of the current-induced response to THz radiation, depending on the metals used and residual doping of graphene. Thermoelectric effect is in the center of all the observations. All-in-all, our work prepares for design of graphene radiation detectors with controllable responsivity.


This work was supported by the FLAG-ERA program (project DeMeGRaS, VR2019-00404, DFG No. GA 501/16-1, and  ANR-19-GRF1-0006), Terahertz Occitanie Platform, and by CNRS through IRP “TeraMIR”. F.T. and S.D.G. thank the support from the IRAP program of the Foundation for Polish Science (grant MAB/2018/9, project CENTERA). This work was performed in part at Myfab Chalmers.

\bibliography{Current-induced_AIP_Adv.bib}

\end{document}


\title{Current-induced enhancement of photo-response in graphene radiation detectors: Supplementary information}
\author{K. Indykiewicz}
\affiliation{Faculty of Electronics, Photonics and Microsystems, Wroc\l{}aw University of Science and Technology, 50-372 Wroc\l{}aw, Poland}
\author{C. Bray }
\affiliation{CNRS/Laboratoire Charles Coulomb (L2C), 34095  Montpellier,  France}
\author{C. Consejo}
\affiliation{CNRS/Laboratoire Charles Coulomb (L2C), 34095  Montpellier,  France}
\author{F. Teppe}
\affiliation{CNRS/Laboratoire Charles Coulomb (L2C), 34095  Montpellier,  France}
\author{S. Danilov}
\affiliation{University of Regensburg, Faculty of Physics, D-93053 Regensburg, Germany}
\author{S.D. Ganichev}
\affiliation{University of Regensburg, Faculty of Physics, D-93053 Regensburg, Germany}
\author{A. Yurgens}
\affiliation{Chalmers University of Technology, SE-412 96 G\"oteborg, Sweden}


\maketitle
\section{Response to microwaves at dc bias}

\begin{figure}[!ht]
\begin{center}
 \includegraphics[width=0.8\textwidth]{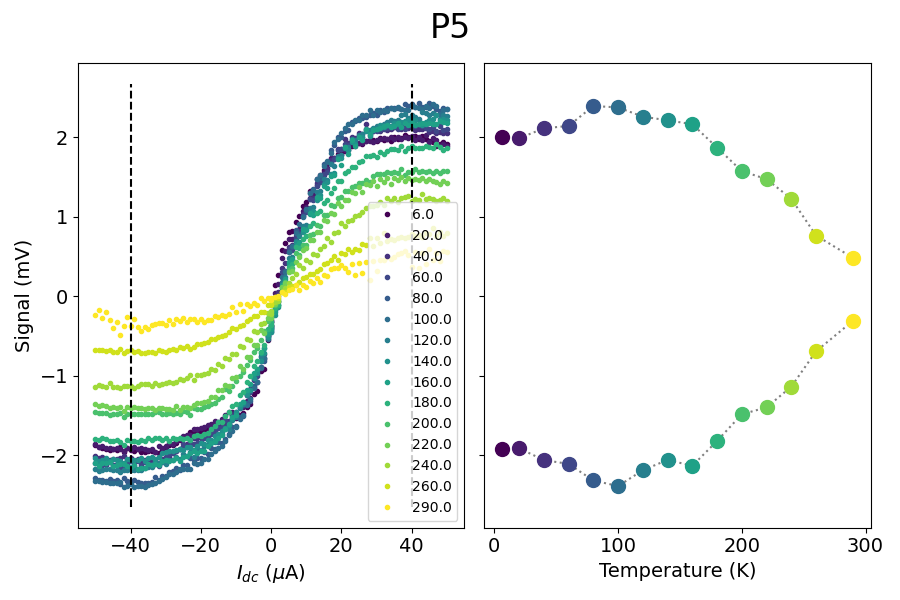}
 \vspace{-5mm}
\end{center}
  \caption{Output signal versus dc current (left) and the temperature dependence of the signal at the two fixed dc currents marked by the vertical dashed lines (right). The top electrodes to graphene were made of Pd in this case.  }
  \label{fig:P5}
\end{figure}

\begin{figure}[!ht]
\begin{center}
 \includegraphics[width=0.8\textwidth]{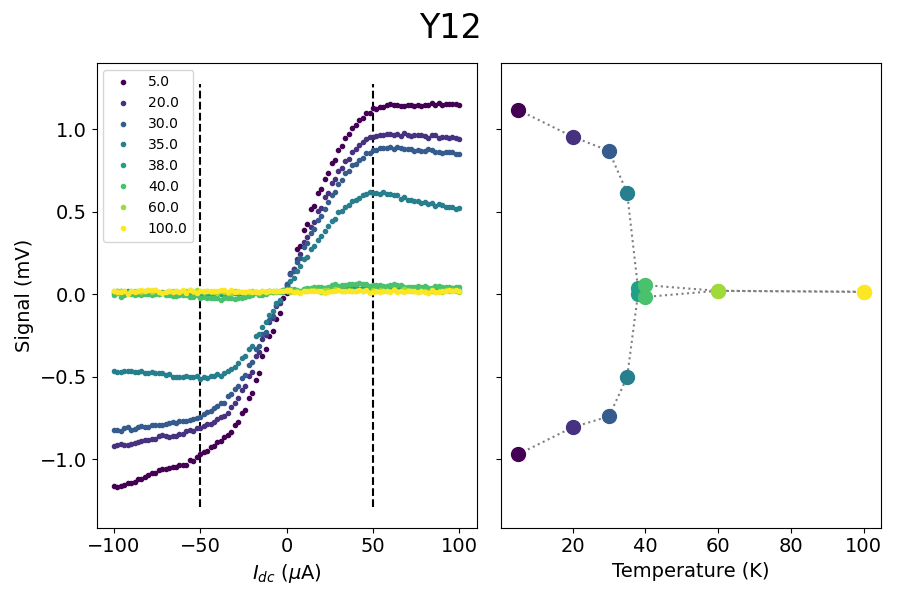}
 \vspace{-5mm}
\end{center}
  \caption{Output signal versus dc current (left) and the temperature dependence of the signal at the two fixed dc currents marked by the vertical dashed lines (right). This sample had a thin layer of $\mathrm{Al_2O_3}$ in between graphene and metal electrode (Pd). Graphene was on top of the electrodes. Note the abrupt disappearance of the signal.}
  \label{fig:Y12}
\end{figure}

Here, we show a couple of other plots demonstrating universality of the discovered effect of dc current, which is largely independent on the materials involved in device fabrication. We observed this effect both in devices with single- and double-layer exfoliated-, and CVD graphene. The CVD graphene, both fabricated in-house- and elsewhere (e.g., Graphenea.com) was always with some density of multi-layer patches, which varied from sample to sample.  Graphene was transferred onto common SiO$_2$/Si- or plastic substrates (EVA/PET, TPX)~\footnote{EVA (ethylene vinyl acetate), PET (polyethylene terephthalate), TPX (polymethylpentene) } and at different fabrication steps, - in the beginning of processing or at the end. In the last case, graphene was transferred onto already patterned electrodes on a SiO$_2$/Si substrate. We also tested to pattern contacts onto freshly grown graphene, before removing the copper foil that was used to grow graphene on. Different metals were used for making the electrodes (antennas) - Pd, Au, Ag, Au/Ge, Al/Al$_2$O$_3$. The effect was observed in full in all cases (see Figs.~2, \ref{fig:P5}, and \ref{fig:Y12}).

\subsection{Fabrication schemes}

\begin{figure}[!ht]
\begin{center}
 \includegraphics[width=0.8\textwidth]{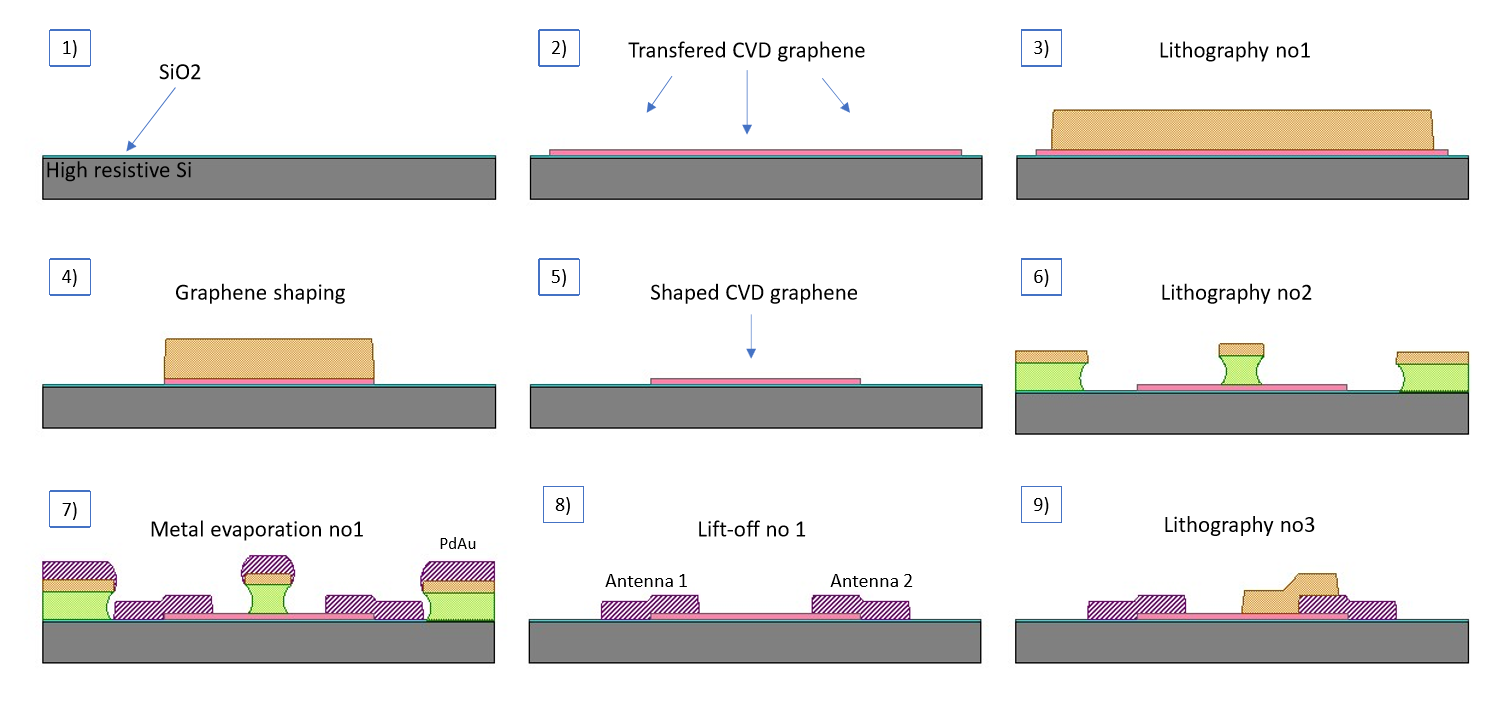}
 \vspace{-5mm}
\end{center}
  \caption{Fabrication scheme for devices with the same metal thin film on two sides of the graphene channel. }
  \label{fig:P5_fab}
\end{figure}

\begin{figure}[!ht]
\begin{center}
 \includegraphics[width=0.8\textwidth]{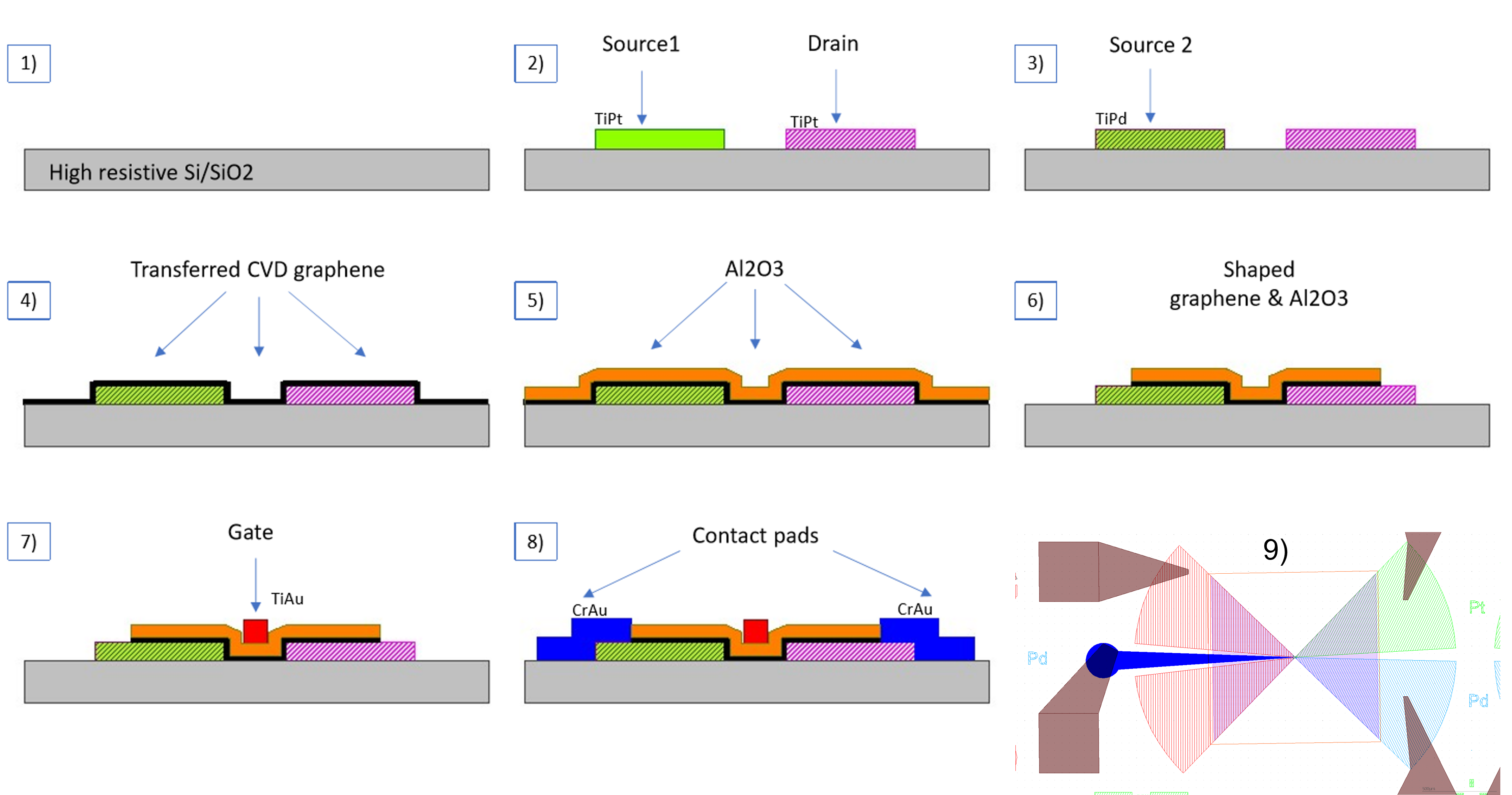}
 \vspace{-5mm}
\end{center}
  \caption{Fabrication scheme for the device with two sources made of dissimilar metals. The device layout is shown in 9). }
  \label{fig:D10_fab}
\end{figure}

Two fabrication schemes for our samples are shown in Figs.~\ref{fig:P5_fab} and \ref{fig:D10_fab}. Since the dc-current effect is present in all the samples, there is no particular step that is crucial and worth emphasizing in these figures.

To see any effect of proximity doping from metal electrodes, several multi-terminal devices were fabricated using metals with dissimilar work functions~\cite{doping-by-metal, cusati_2017}. Pt and Pd are the metals, which are expected to dope graphene with charges of opposite signs~\cite{doping-by-metal, cusati_2017}.  No significant difference between the samples was however observed, only a small shift of the signal-vs-dc-current curves corresponding to a non-zero signal at the zero dc current. This is understandable because dissimilar metals (and graphene doping level under them) would also break the symmetry of the device even without dc current.

Following quite a naive idea to increase the voltage drop between the metal and graphene at a given dc current (self-gating, see below), a less conducting Ge layer between Au and graphene was also tried. Although the maximum signal was noted to increase somewhat, this did not dramatically change the detector performance. This can be understood from the schematic cross-section picture of the devices (see Fig.~1c). The current-driven self-gating is ultimately decided by the vacuum gap between graphene and the closest to graphene material, i.e., by the parallel-plate capacitance with graphene as one of the plates. The voltage drop across the material layer itself has no major impact on the doping because the layer is literally farther away from graphene.

The additional \emph{p-n} junction in the middle of the graphene channel had only a minor effect on the resulting characteristics. To create such a junction, we covered one half of the channel by photoresist and applied some chemicals at hand~\footnote{A water-based mixture commonly used to inhibit accumulation of static charges on surfaces; see www.aclstaticide.com for more details. That it produced a negative doping to graphene was found in independent experiments.} that induced negative doping to another (open) half of the channel. Since transferred CVD graphene is normally \emph{p}-doped, protected graphene will remain positively doped, while the open graphene will become \emph{n}-doped. A \emph{p-n} junction could thus be made in the middle of the graphene channel.

\section{Raman spectra}

\begin{figure}[!ht]
	\begin{center}
		\includegraphics[width=0.8\textwidth]{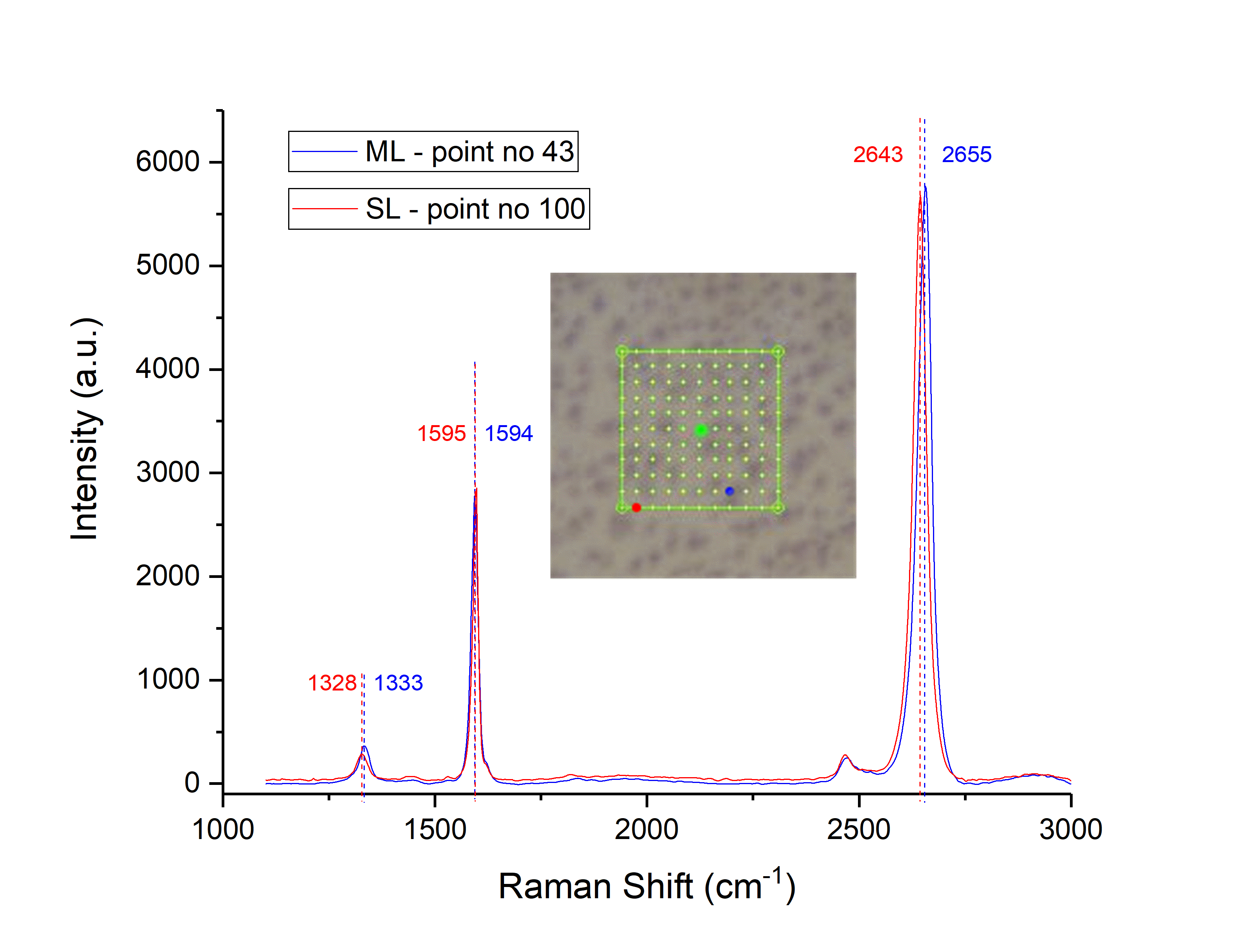}
		\vspace{-5mm}
	\end{center}
	\caption{Raman spectrum of CVD graphene transferred to SiO$_2$/Si substrate, with the characteristic peaks labeled by their Raman-shift energies and vertical lines (red - single-layer locations, SL; blue - multilayer ones, ML).  The inset shows an optical image of the graphene. Numerous multilayer patches are seen. The Raman-mapping locations are marked by dots within the green frame $40\times 40$~$\mu$m$^2$ large. Note that the 2D peak is symmetric both for SL and ML regions. This indicates that the graphene layers are twisted relative to each other.   }
	\label{fig:Raman}
\end{figure}

Raman-spectra microscope (Horiba XploRA, 638 nm excitation laser) was routinely used to characterize our CVD- and exfoliated graphene (see Fig.~\ref{fig:Raman}). In the case graphene was still on copper- or transferred to EVA/PET foil, Raman spectra were "contaminated" by either strong background from copper or several additional peaks from EVA and PET. The latter overlapped with e.g., D and G peaks of graphene, which often obstructed quantitative analysis of graphene quality.   

Our CVD graphene always had multilayer patches (see the inset in Fig.~\ref{fig:Raman}). However, 2D peak  retained its symmetrical shape in the majority of cases. This indicated that the individual graphene layers in the patches were much likely twisted relative to each other~\cite{ferrari_raman_2013}. 

\section{Modeling}

The charge-density (doping) profile $c(x)$ is approximated by: $c(x) = c_p-(c_n + c_p)[F(-x/\delta + s/2/\delta) + F(x/\delta + s/2/\delta)]$, where $F(u)=1/[1+\exp(u)]$ is the Fermi function, $c_n$ and $c_p$ are the maximum doping levels of $n$- and $p$ regions, respectively; $s$ is the separation between the electrodes, and $\delta$ determines the smearing of the profile due to e.g., fringing of the electric field at the contact edges. For brevity of equations, $c$ has sign corresponding to the sign of charge carriers.

The  electrical conductance $\sigma$ changes very little with temperature and therefore is taken to depend on $x$ only: $\sigma (c) =\mu |e| \sqrt{c^2 + c_0^2}$, where $\mu$  is the charge-carrier mobility, $e$ is the elementary charge, and $c_0$ is the residual charge density.

The Seebeck coefficient $S$ in graphene is assumed to obey Mott's equation in the whole temperature range.
\begin{equation}
	S = \frac{\pi ^2}{3} \frac{k_B}{|e|} k_B T \frac{\partial\ln{\sigma }}{\partial c} \frac{\partial c}{\partial E_F}
	\label{eq:Mott}
\end{equation}
\noindent where $k_B$ is the Boltzmann constant and $E_F$ is the Fermi energy.

Numerically solving Eqs.~1 and 2  gives  $T_e(x)$, $T_{ph}(x)$, the TEP voltage, and the total Joule dissipation for any bias current $j$. The current in real detectors will be a sum of the dc- and ac components: $j(t)=j_{dc} + j_0 \sin(\omega t)$. For $\omega\ll 2\pi/\tau$, where $\tau<50$~fs is the electron-heating time \cite{Koppens_fs_nnano}, the responsivity $\Re$ can be found by averaging the voltage and Joule power over one period of the ac bias:
\begin{eqnarray}
	V_\mathrm{TEP} &=& \Big\langle \int_0^l S(x)\nabla T_e(x) \mathrm{d}x \Big\rangle\\
	P_\mathrm{tot} &=& \Big\langle j^2 \int_0^l \frac{w\mathrm{d}x}{\sigma(x)} \Big\rangle\\
	\Re            &=& V_\mathrm{TEP}/P_\mathrm{tot}
\end{eqnarray}

\subsection{The temperature profile}

\begin{figure}[!ht]
\begin{center}
 \includegraphics[width=0.5\textwidth]{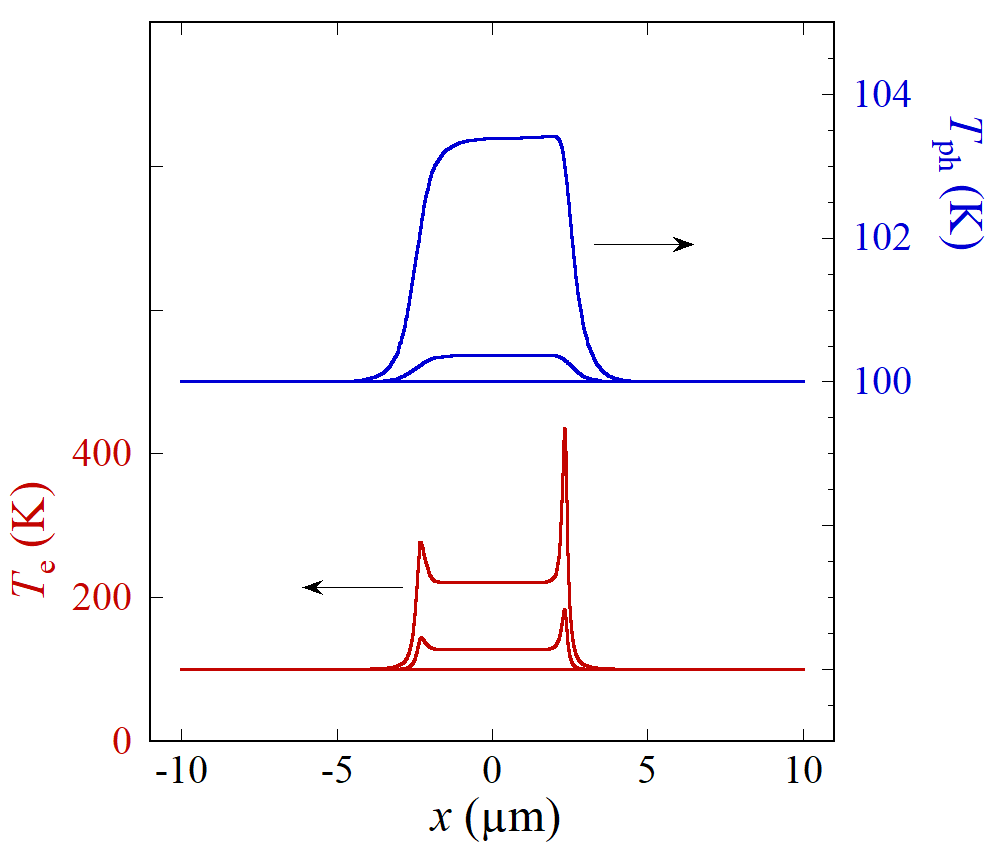}
 \vspace{-5mm}
\end{center}
  \caption{The electron ($T_e$)- and phonon ($T_{ph}$) temperature profiles at two dc currents, 15 and 45 A/m and ambient temperature $T_0=100$~K. }
  \label{fig:T(x)}
\end{figure}

The electron ($T_e$)- and phonon ($T_{ph}$) temperature profiles at two dc currents, 15 and 45 A/m are shown in Fig.~\ref{fig:T(x)}. It is seen that both $T_e$ and $T_{ph}$ are asymmetric due to Peltier effect and $T_e$ is the highest at the regions of \emph{p-n} junctions, with the highest Joule heating. The asymmetry is the main reason for the response signal in our devices.

\subsection{Single p-n junction}

It is known that a p-n junction in the middle of the graphene channel will lead to a rectification of the incoming radiation, representing foundation for building radiation sensors.  In contrast to semiconductor diodes, this rectification stems from the Joule heating and thermoelectric effect~\cite{TEP_2014,Skoblin_APL,me_Sensors}. The question is if this rectification can be further improved by dc current. Indeed, the dc current will increase the overall Joule heating and electronic temperature. Consequently, the temperature gradients are anticipated to increase as well, resulting in a larger thermoelectric voltage (signal). The Peltier effect is expected to induce an asymmetry of the signal with respect to direction of the dc current.

\begin{figure}[!ht]
\begin{center}
 \includegraphics[width=0.5\textwidth]{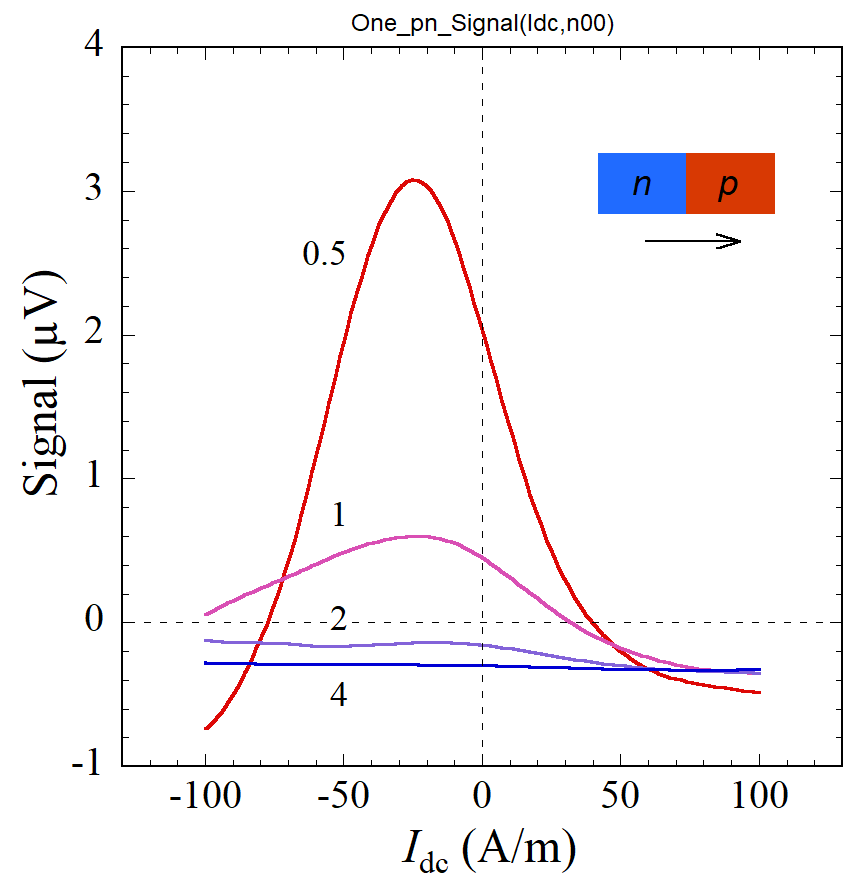}
 \vspace{-5mm}
\end{center}
  \caption{Simulations of the output signal as a function of the dc current at different residual charge densities indicated near the curves (in units $10^{12}\ \mathrm{cm}^{-2}$). The inset schematically shows the positive direction of the current, which corresponds to Peltier cooling. }
  \label{fig:One_pn}
\end{figure}

To answer this question, we repeated the numerical simulations of Ref.~\cite{me_Sensors}, now with the dc-current component included. The results of the simulations are shown in Fig.~\ref{fig:One_pn}. For small dc current and residual charge density, the rectified signal increases (decreases) for the current direction corresponding to heating (cooling) of the \emph{p-n} junction. Further increasing the current absolute value results in decrease of the signal even for the negative direction. The signal becomes more or less independent of the current for high $n_{00}$. All this shows that the dc current is largely useless for increasing the signal in the case of isolated \emph{p-n} junctions.

\subsection{Perpendicular polarization}

Here, we demonstrate that dc current results in an increased response  also for the perpendicular polarization of THz radiation (relative to the line connecting two halves of the antenna). This situation is modeled by letting the ac current to only flow in the open part of the graphene channel and assuming that the graphene parts under the metal contacts are well shielded from the radiation. The dc current flows between the source and drain, as usual. 

\begin{figure}[!ht]
	\begin{center}
		\includegraphics[width=0.8\textwidth]{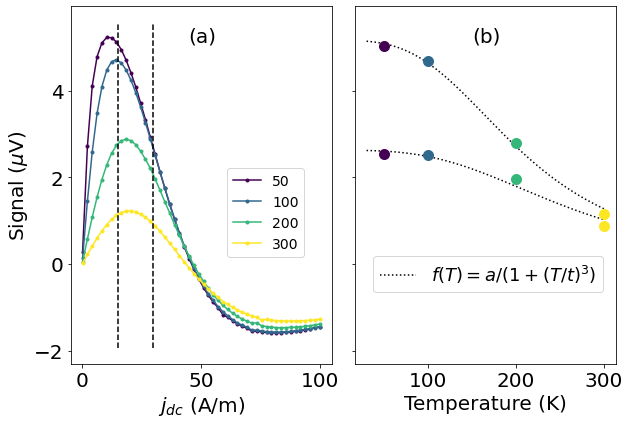}
		\vspace{-5mm}
	\end{center}
	\caption{The simulated response signal versus dc current at different temperatures. The following parameters were used: ($c_{00}$, $c_p$, $c_n$) = (0.5, 1, -5)$\times 10^{12}$~cm$^{-2}$, $\lambda_j=1\ \mathrm{\mu m}$, $j_{ac}=1$~A/m (see also Fig.~1b). $j_{ac}(x)=0$ everywhere except the open part of graphene. The inset shows the temperature dependence of the maximum signal.  }
	\label{fig:static_n_signal_perp}
\end{figure}

The results are shown in Fig.~\ref{fig:static_n_signal_perp}. The curves are largely the same as in Fig.~5, with just somewhat smaller maximum values because of smaller Joule heating from the ac current.  In the usual parallel-polarization case, because of the current crowding, the ac current flows in only a little  larger areas ($\sim 2w_0 \lambda_j$). This explains the weak dependence of the effect on the THz-light polarization for a large impedance mismatch with antennas.

\subsection{Self-gating under contacts}
From Fig.~1d it is straightforward to realize that the current $j_r(x)$ flowing through the contact resistance $r$ gives rise to a voltage drop $v_l(x)$ across the parallel-plate contact capacitance $C$, thereby creating additional charges in the graphene. Consequently, the (local) graphene resistance, $R=\mu^{-1} (C^2V_l^2+c_0^2e_0^2)^{-0.5}$,  changes as well, which reciprocally affects the current $j_r(x)$. This kind of back-action or self-dependency can introduce a significant non-linearity in the contact current-voltage  characteristics (I-V's), to be distinguished from non-linear \emph{tunneling} I-V's, which reflect variations of the density of states with energy.  

The importance of the self-gating effect in the top contacts to graphene stems from the huge, never mind how "leaky", areal capacitance of the contacts, $c_s=\epsilon_0/(0.34\ \mathrm{nm})\approx 26$~mF/m$^2$.  It is clear that $c_s$ is on par with the quantum capacitance. For $v_l$ of just 100~mV, the additional charge density in graphene is $c_s v_l/e_0=1.6\times 10^{16}$~1/m$^2$, i.e., a substantial part of the usual overall doping in graphene.

\begin{figure}[!ht]
		\begin{center}
			\includegraphics[width=0.5\textwidth]{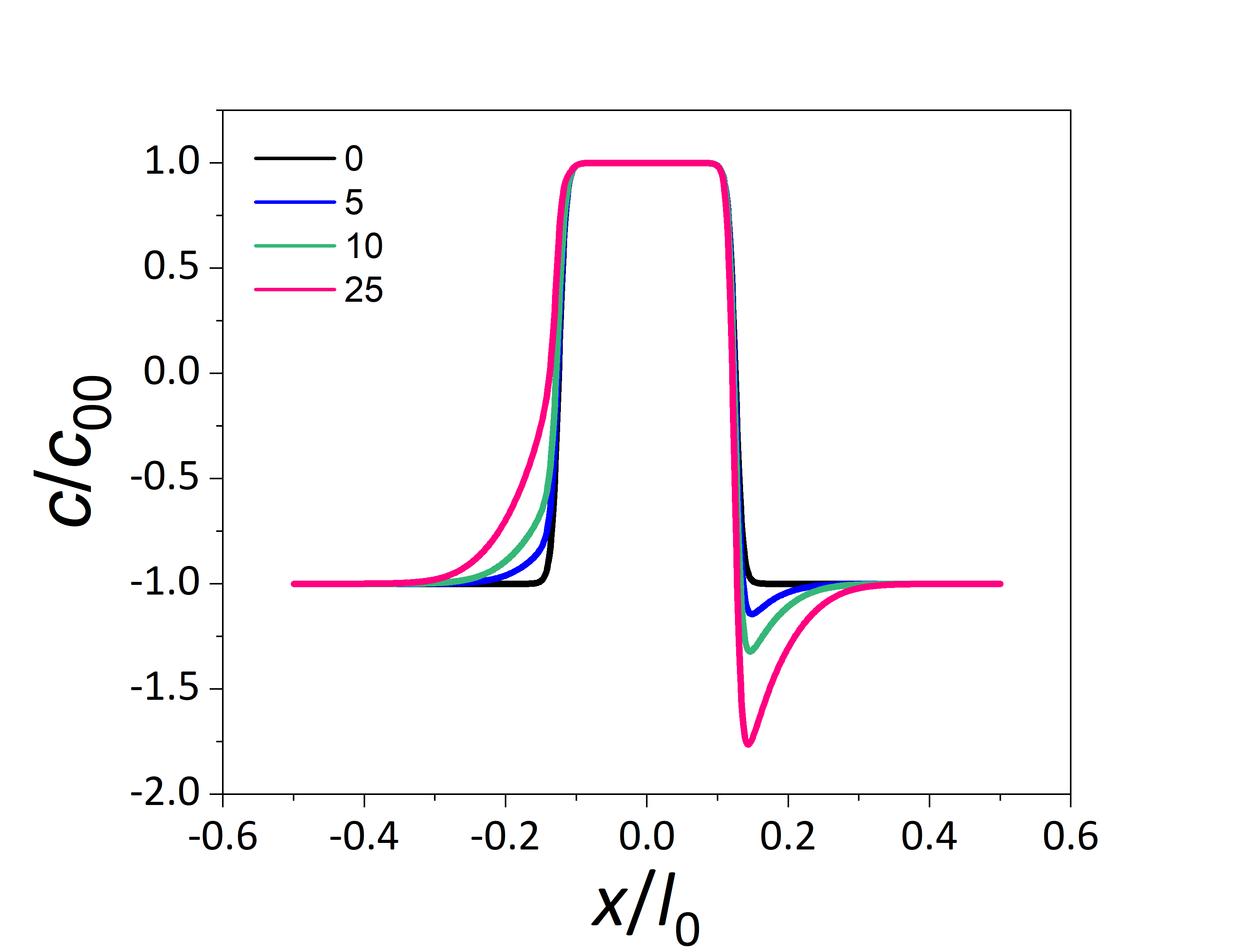}
			\vspace{-5mm}
		\end{center}
		\caption{ The self-doping profiles for different dc currents in [A/m]. The current-crowding length is 1~$\mu$m, $s=5$~$\mu$m, $l_0=20$~$\mu$m, and   $c_{00} = c_p = -c_n = 10^{12}\ \mathrm{cm}^{-2}$.  }
		\label{fig:self-doping_profile}
	\end{figure}

	Fig.~\ref{fig:self-doping_profile} shows the combined proximity plus self-gating doping profile for a set of parameters, also taking into account the current crowding (see Figs.~1c and 1d). The initially symmetric profile at $j=0$ gets significantly skewed with increasing $j$.
	
Of course, the contacts in real devices are never perfect. There is much likely a mix of both the plane top contacts (representing vacuum tunneling) and edge contacts (involving covalent bonding between carbon- and metal atoms). The latter contacts can take place through, e.g., numerous local defects in graphene and also because the contacts almost always cross the edge of a patterned graphene strip. No wonder that the self-gating non-linearity described above has never been unambiguously singled out in experiments.

\begin{figure}[!ht]
	\begin{center}
		\includegraphics[width=0.8\textwidth]{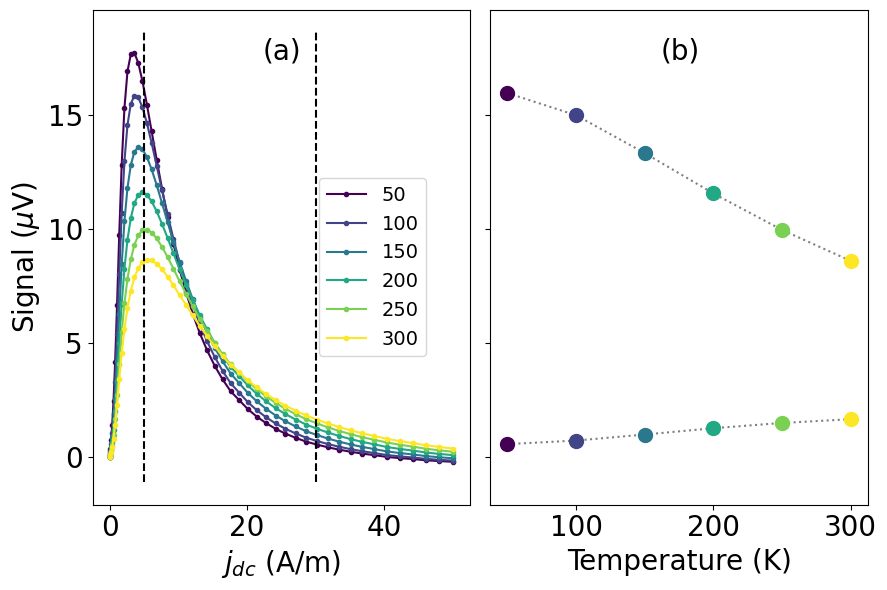}
		\vspace{-5mm}
	\end{center}
	\caption{The simulated response signal versus dc current at different temperatures, taking into account the self-gating effect and quantum capacitance. The following parameters were used: ($c_{00}$, $c_p$, $c_n$) = (0.5, 0.01, -0.01)$\times 10^{12}$~cm$^{-2}$, $\lambda_j=1\ \mathrm{\mu m}$, $j_{ac}=1$~A/m.  The inset shows the temperature dependence of the maximum signal.  }
	\label{fig:self-gating_n_signal}
\end{figure}

Here, we demonstrate how the self-gating would alter the dc-current-induced photo-response in symmetrical graphene structures. We assume that the initial (static) doping distribution in graphene is still decided by the proximity to metal contacts and initial doping from lithography processing. In addition, now we let the doping be affected by the local electric field as well. The current crowding and quantum capacitance are taken into account. The results are shown in Fig.~\ref{fig:self-gating_n_signal}. With except for more "noise" at high current, the curves are essentially the same as in Fig.~5, again emphasizing that it is \emph{p-n} junctions that contribute most to detection of the THz radiation.

\bibliography{Suppl_Current-induced_AIP_Adv.bib}